\documentclass[twocolumn]{svjour3}          
\smartqed  
\usepackage{graphicx}
 \usepackage{mathptmx}      
 \usepackage[utf8]{inputenc}
 \usepackage[T1]{fontenc}
 \usepackage[english]{babel}
 \usepackage{amsmath}
 \usepackage{amssymb}
 
\begin{document}

\title{Shot noise in ultrathin superconducting wires}


\author{Andrew G. Semenov    \and
        Andrei D. Zaikin
}


\institute{A.G. Semenov \at
           1) I.E.Tamm Department of Theoretical Physics, P.N.Lebedev Physics Institute, 119991 Moscow, Russia\\
           2) National Research University Higher School of Economics, 101000 Moscow, Russia\\
              \email{semenov@lpi.ru}           
           \and
           A.D. Zaikin \at
           1) Institute of Nanotechnology, Karlsruhe Institute of Technology (KIT), 76021 Karlsruhe, Germany\\
           2) I.E.Tamm Department of Theoretical Physics, P.N.Lebedev Physics Institute, 119991 Moscow, Russia\\
              \email{andrei.zaikin@kit.edu}           
}

\date{Received: date / Accepted: date}

\maketitle

\begin{abstract}
Quantum phase slips (QPS) may produce non-equilibrium voltage fluctuations in current-biased superconducting nanowires. Making use of the Keldysh technique and employing the phase-charge duality arguments we investigate such fluctuations within the four-point measurement scheme and demonstrate that shot noise of the voltage detected in such nanowires may essentially depend on the particular measurement setup. In long wires the shot noise power decreases with increasing frequency $\Omega$ and vanishes beyond a threshold value of $\Omega$ at $T \to 0$.
\end{abstract}

\section{Introduction}
\label{intro}
Electric current can flow through a superconducting material without any resistance. This is perhaps the most fundamental property of any bulk superconductor which properties are usually well described by means of the
standard mean field theory approach. The situation changes, however, provided (some) superconductor dimensions become sufficiently small. In this case thermal and/or quantum fluctuations may set in and
the system properties may qualitatively differ from those of bulk superconducting structures.

In what follows we will specifically address fluctuation effects in ultrathin superconducting wires.
In the low temperature limit thermal fluctuations in such wires are of little importance and their behaviour is dominated by
the quantum phase slippage process \cite{AGZ} which causes local temporal suppression of the superconducting order
parameter $\Delta=|\Delta|e^{i\varphi}$ inside the wire. Each such quantum phase slip (QPS) event implies the
net phase jump by $\delta \varphi =\pm 2\pi$ accompanied by a voltage pulse $\delta V=\dot{\varphi}/2e$ as well as
tunneling of one magnetic flux quantum $\Phi_0\equiv \pi/e =\int |\delta V(t)|dt$ across the wire normally to its axis.
Different QPS events can be viewed as logarithmically interacting quantum particles \cite{ZGOZ} forming a 2d gas in space-time
characterized by an effective fugacity proportional to the QPS tunneling amplitude per unit wire length \cite{GZQPS}
\begin{equation}
\gamma_{QPS} \sim (g_\xi\Delta_0/\xi)\exp (-ag_\xi), \quad a \sim 1.
\label{gQPS}
\end{equation}
Here $g_\xi =2\pi\sigma_N s/(e^2\xi) \gg 1$ is the dimensionless normal state conductance of the
wire segment of length equal to the coherence length $\xi$, $\Delta_0$ is the mean field order parameter value, $\sigma_N$ and $s$ are respectively the wire Drude conductance and the wire cross section.

In the zero temperature limit a quantum phase transition occurs in long superconducting wires \cite{ZGOZ} controlled by the parameter
$\lambda \propto \sqrt{s}$ to be specified below. In thinnest wires with $\lambda < 2$ superconductivity is totally washed out by quantum fluctuations. Such wires may go insulating
at $T=0$. In comparatively thicker wires with $\lambda > 2$ quantum fluctuations are less pronounced, the wire resistance $R$ decreases with $T$ and takes the form \cite{ZGOZ}
\begin{equation}
\label{Rdiff}
R=\frac{d\langle \hat V\rangle}{dI}
\propto
\begin{cases}
\gamma_{QPS}^2T^{2\lambda -3}, & T\gg \Phi_0I,
\\
\gamma_{QPS}^2I^{2\lambda -3}, & T\ll \Phi_0I.
\end{cases}
\end{equation}
Hence, the wire non-linear resistance remains non-zero down to lowest temperatures, just as it was observed in a number of experiments  \cite{BT,Lau,Zgi08,liege}.

The result (\ref{Rdiff}) combined with the fluctuation-dissipation theorem (FDT) implies that
equilibrium voltage fluctuations develop in superconducting nanowires in the
presence of quantum phase slips. One can also proceed beyond FDT and
demonstrate \cite{SZ16a,SZ16b} that quantum phase slips may generate {\it non-equilibrium} voltage fluctuations in ultrathin superconducting wires. Such fluctuations are associated with the process of quantum tunneling of magnetic flux quanta $\Phi_0$ and turn out to obey Poisson statistics. The
QPS-induced shot noise in such wires is characterized by a highly non-trivial dependence of its power spectrum on temperature, frequency and external current.

Note that in Refs. \cite{SZ16a,SZ16b} we addressed a specific noise measurement scheme with a voltage detector placed at one end of a
superconducting nanowire while its opposite end was considered grounded. The main goal of our present work is to demonstrate that
quantum shot noise of the voltage in such nanowires may essentially depend on the particular measurement setup. Below we will
evaluate QPS-induced voltage noise within the four-point measurement scheme involving two voltage detectors and compare our results with derived earlier
\cite{SZ16a,SZ16b}.

\section{System setup and basic Hamiltonian}

Let us consider the system depicted in Fig. 1. It consists of a superconducting nanowire attached to a current source $I$ and two voltage probes located in the points $x_1$ and $x_2$. The wire contains a thinner segment of length $L$ region where quantum phase slips can occur with the amplitude (\ref{gQPS}).
\begin{figure}
  \includegraphics[width=\columnwidth]{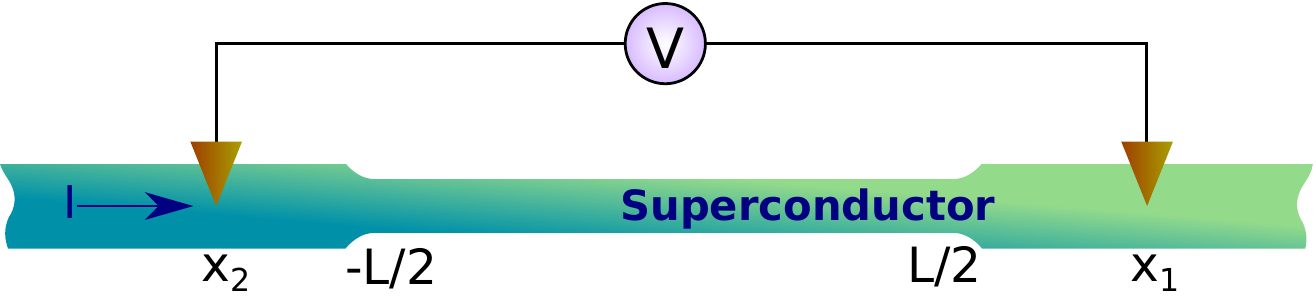}
  \caption{The system under consideration.}
\end{figure}

In order to proceed with our analysis of voltage fluctuations we will make use of the duality arguments \cite{SZ16a,SZ16b}. The effective dual low-energy Hamiltonian of our system has the form
\begin{equation}
\hat H_{\rm wire} = \hat H_{TL}+ \hat H_{QPS}.
\label{Hamw}
\end{equation}
The term
\begin{equation}
\hat H_{TL}=\int\limits_{-\infty}^{\infty} dx\left(\frac{1}{2\mathcal{L}_{\rm kin}}\left(\hat\Phi(x)+\mathcal{L}_{\rm kin} I\right)^2+\frac{e^2(\nabla\hat\chi(x))^2}{2C_{\rm w}\Phi_0^2}
\right)
\label{Htl}
\end{equation}
defines the wire Hamiltonian in the absence of quantum phase slips. It describes an effective transmission line in terms of two canonically conjugate operators $\hat \Phi (x)$ and $\hat \chi (x)$ obeying
the commutation relation
\begin{equation}
[\hat \Phi (x),\hat \chi (x')]=-i\Phi_0\delta (x-x').
\label{commrel}
\end{equation}
Here and below $C_{\rm w}$ denotes geometric capacitance per unit wire length and $\mathcal{L}_{\rm kin}=1/(\pi\sigma_N\Delta_0 s)$ is the wire kinetic
inductance (times length).  In Eqs. (\ref{Htl}), (\ref{commrel}) and below $\hat \Phi (x)$ is the magnetic flux operator while the quantum field $\hat \chi (x)$ is proportional to the total charge $\hat q(x)$ that has passed through
the point $x$ up to the some time moment $t$, i.e. $\hat q(x)=-\hat \chi (x)/\Phi_0$. Hence, the local charge density $\hat \rho (x)$ and the phase difference between the
two wire points $x_1$ and $x_2$ can be defined as
\begin{equation}
\hat \rho (x)=\nabla\hat \chi (x)/\Phi_0,
\end{equation}
\begin{equation}
\hat\varphi(x_1)-\hat\varphi(x_2)=2e\mathcal{L}_{\rm kin} I(x_1-x_2)+2e\int\limits_{x_2}^{x_1}dx\hat\Phi(x).
\end{equation}
Employing the expression for the local charge density (or, alternatively, the Josephson relation) it is easy to recover
the expression for the operator corresponding to the voltage difference between the points $x_1$ and $x_2$. It reads
\begin{equation}
\hat V=\frac{1}{\Phi_0 C_{\rm w}}\left(\nabla\hat\chi(x_1)-\nabla\hat\chi(x_2)\right).
\end{equation}

The term $\hat H_{QPS}$ in Eq. (\ref{Hamw}) accounts for QPS effects and has the form \cite{SZ13}
\begin{equation}
\hat H_{QPS}=-\gamma_{QPS}\int_{-L/2}^{L/2}dx \cos(\hat \chi(x)).
\label{HQPS}
\end{equation}
It is easy to note that this term is exactly dual to that describing the Josephson coupling energy in spatially extended Josephson tunnel junctions.

\section{Keldysh technique and perturbation theory}

In order to investigate QPS-induced voltage fluctuations in our system we will employ the Keldysh path integral technique. We routinely
define the variables of interest on the forward and backward time parts of the Keldysh contour, $\chi_{F,B}(x,t)$, and introduce the ``classical'' and ``quantum'' variables, respectively
\begin{equation}
\chi_+ (x,t)= (\chi_F (x,t)+\chi_B (x,t))/2
\end{equation}
and
\begin{equation}
\chi_- (x,t)= \chi_F (x,t)-\chi_B (x,t).
\end{equation}
Any general correlator of voltages can be represented in the form \cite{SZ16b}
\begin{equation}
\langle V(t_1)V(t_2)...V(t_n)\rangle =\left\langle V_{+}(t_1) V_{+}(t_2)... V_{+}(t_n) e^{iS_{QPS}}\right\rangle_0,
\label{VVn}
\end{equation}
where
\begin{equation}
V_+(t)=\frac{1}{\Phi_0 C_{\rm w}}\left(\nabla\chi_+(x_1,t)-\nabla\chi_+(x_2,t)\right),
\end{equation}
\begin{equation}
 S_{QPS}=-2 \gamma_{QPS}\int dt\int\limits_{-L/2}^{L/2} dx\sin(\chi_{+}(x,t))\sin (\chi_-(x,t)/2)
\end{equation}
and
\begin{equation}
\langle ...\rangle_0 =\int \mathcal D^2\chi (x,t) (...) e^{iS_0[\chi ]}
\end{equation}
indicates averaging with the effective action $S_0[\chi ]$  corresponding to the Hamiltonian $\hat H_{TL}$.
It is important to emphasize that Eq. (\ref{VVn}) defines the symmetrized voltage correlators. E.g.,
for the voltage-voltage correlator one has
\begin{equation}
\langle V(t_1)V(t_2)\rangle =\frac12\langle\hat V(t_1)\hat V(t_2)+\hat V(t_2)\hat V(t_1)\rangle,
\label{VV2}
\end{equation}

In order to evaluate formally exact expressions for the voltage correlators (\ref{VVn}), (\ref{VV2})
one employ the perturbation theory expanding these expressions in powers of the QPS amplitude $\gamma_{QPS}$. It is easy to verify that
the first order terms in this expansion vanish and one should proceed up to the second order in $\gamma_{QPS}$. Due to the quadratic structure of
the action $S_0$ the result is expressed in terms of the average
$\langle\chi_+(x,t)\rangle_0=\Phi_0It$
and the Green functions
\begin{multline}
G^K(x-x',t-t')=-i\langle \chi_{+}(x,t)\chi_{+}(x',t')\rangle_0 \\+i \langle \chi_{+}(x,t)\rangle_0\langle\chi_{+}(x',t')\rangle_0 ,
\end{multline}
\begin{equation}
G^R(x-x',t-t')=-i\langle \chi_{+}(x,t)\chi_{-}(x',t')\rangle_0.
\label{GRK}
\end{equation}
The Keldysh function $G^K$ can also be expressed in the form
\begin{equation}
 G^K(x,\omega)=\frac12\coth\left(\frac{\omega}{2T}\right)\left(G^R(x,\omega)-G^R(x,-\omega)\right).
\end{equation}
A simple calculation allows to explicitly evaluate the retarded Green function, which in our case reads
\begin{equation}
G^R(x,\omega)=-\frac{2\pi i\lambda}{\omega+i0}e^{i\frac{\omega|x-x'|}{v}}.
\end{equation}
Here  $v=1/\sqrt{{\mathcal L}_{\rm kin}C_{\rm w}}$ is the plasmon velocity \cite{ms}
and the parameter $\lambda$  is defined as $\lambda=R_Q/(2Z_{\rm w})$, where $R_Q =\pi/(2e^2)$ is the "superconducting" quantum resistance unit
and $Z_{\rm w}=\sqrt{\mathcal{L}_{\rm kin}/C_{\rm w}}$ being the wire impedance.

\section{Voltage noise}

The above expressions allow to directly evaluate voltage correlators perturbatively in  $\gamma_{QPS}$. In the case of the four-point measurement scheme
of Fig. 1 the calculation is similar to one already carried out for the two-point measurements \cite{SZ16a,SZ16b}. Therefore we can directly proceed to
our final results. Evaluating the first moment of the voltage operator $\langle\hat V\rangle$ we again reproduce the results \cite{ZGOZ,SZ16a} which yield
Eq. (\ref{Rdiff}). For the voltage noise power spectrum $S_{\Omega}$ we obtain
\begin{equation}
 S_{\Omega}=\int dt e^{i\Omega t}\langle V(t)V(0)\rangle =S_{\Omega}^{(0)}+S_{\Omega}^{QPS},
\label{VVS}
\end{equation}
where the term $S_{\Omega}^{(0)}$ describes equilibrium voltage noise in the absence of QPS (which is of no interest for us here) and
\begin{multline}
S_{\Omega}^{QPS}=\frac{\gamma_{QPS}^2e^2\coth\left(\frac{\Omega}{2T}\right)}{4\pi^2C_{\rm w}^2}\int\frac{dk}{2\pi}\mathcal M_k(\Omega)\big(
 P_k(\pi I/e)\\-\bar P_k(-\pi I/e)+P_k(-\pi I/e)-\bar P_k(\pi I/e)\big)
\\+ \frac{\gamma_{QPS}^2e^2\coth\left(\frac{\Omega}{2T}\right)}{4\pi^2C_{\rm w}^2}\int\frac{dk}{2\pi}\mathcal S_k(\Omega)\mathcal S_{-k}(\Omega)\big(
 \bar P_k(-\Omega-\pi I/e)\\- P_k(\Omega+\pi I/e)+\bar P_k(-\Omega+\pi I/e)- P_k(-\Omega-\pi I/e)\big)
 \\+\frac{\gamma_{QPS}^2e^2\left(\coth\left(\frac{\Omega+\pi I/e}{2T}\right)-\coth\left(\frac{\Omega}{2T}\right)\right)}{4\pi^2C^2_{\rm w}}\int\frac{dk}{2\pi}\mathcal S_k(\Omega)\\\times\mathcal S_{-k}(-\Omega)\big(
 P_k(\Omega+\pi I/e)+\bar P_k(\Omega+\pi I/e)\\-P_k(-\Omega-\pi I/e)-\bar P_k(-\Omega-\pi I/e)\big)
+\big(\Omega\to-\Omega\big)
\label{SQPS}
\end{multline}
is the voltage noise power spectrum generated by quantum phase slips. Eq. (\ref{SQPS}) contains the function
\begin{equation}
 \label{P}
 P_{k}(\omega)=\int d x e^{ikx}\int\limits_{0}^\infty dt e^{i\omega t}e^{iG^K(x,t)-iG^K(0,0)+\frac i2 G^R(x,t)}
\end{equation}
and geometric form-factors $\mathcal M_k(\Omega)$ and $\mathcal S_k(\Omega)$ which explicitly depend on $x_1$ and $x_2$. E.g., for $x_1=L/2$ and $x_2=-L/2$ we get
\begin{multline}
\mathcal M_k(\Omega)=(4\pi\lambda)^2e^{\frac{i\Omega L}{v}}\frac{\sin\left(\frac{kL}{2}\right)}{vk}\Biggl(\frac{2\sin\left(\frac{kL}{2}\right)}{vk}\\+\frac{\sin\left(\frac{(2\Omega+vk)L}{2v}\right)}{2\Omega+vk}+\frac{\sin\left(\frac{(2\Omega-vk)L}{2v}\right)}{2\Omega-vk}\Biggr),
\end{multline}
\begin{equation}
\mathcal S_k(\Omega)=4\pi\lambda e^{\frac{i\Omega L}{2v}}\Biggl(\frac{\sin\left(\frac{(\Omega+vk)L}{2v}\right)}{\Omega+vk}+\frac{\sin\left(\frac{(\Omega-vk)L}{2v}\right)}{\Omega-vk}\Biggr).
\end{equation}
We observe that these form-factors oscillate as functions of $\Omega$. Such oscillating behavior stems from the interference effect at the boundaries of a thinner wire segment making the result for the shot noise
in general substantially different as compared to that measured by means of a setup with one voltage detector \cite{SZ16a,SZ16b}. In the long wire limit and for $\Omega\gg v/L$ one has
\begin{equation}
\mathcal M_k(\Omega)\approx \pi L\frac{(4\pi\lambda)^2}{v^2} e^{\frac{i\Omega L}{v}}\delta(k),
\label{Mk}
\end{equation}
\begin{multline}
\mathcal S_k(\Omega)\mathcal S_{-k}(\Omega)\approx\frac{(4\pi\lambda)^2}{v^2} e^{\frac{i\Omega L}{v}}\biggl(\frac{\pi L}{2}\delta\left(k+\frac{\Omega}{v}\right)\\+\frac{\pi L}{2}\delta\left(k-\frac{\Omega}{v}\right)\biggr),
\label{Sk}
\end{multline}
\begin{multline}
\mathcal S_k(\Omega)\mathcal S_{-k}(-\Omega)\approx\frac{(4\pi\lambda)^2}{v^2} \biggl(\frac{\pi L}{2}\delta\left(k+\frac{\Omega}{v}\right)\\+\frac{\pi L}{2}\delta\left(k-\frac{\Omega}{v}\right)\biggr).
\label{SkSk}
\end{multline}
Neglecting the contributions (\ref{Mk}) and (\ref{Sk}) containing fast oscillating factors $e^{\frac{i\Omega L}{v}}$ and combining the remaining term (\ref{SkSk}) with Eq. (\ref{SQPS}), we obtain
\begin{multline}
S_{\Omega}^{QPS}=\frac{L\pi^2\gamma_{QPS}^2}{8e^2}\frac{\sinh\left(\frac{\Phi_0 I}{2T}\right)\varsigma\left(\frac{\Phi_0 I}{2}\right)}{\sinh\left(\frac{\Omega}{2T}\right)}\\
\times \biggl(\varsigma\left(\frac{\Phi_0 I}{2}-\Omega\right)
-\varsigma\left(\frac{\Phi_0 I}{2}+\Omega\right)\biggr).
\label{Sa3}
\end{multline}
where
\begin{equation}
\varsigma(\omega)=\tau_0^\lambda (2\pi T)^{\lambda -1}\frac{\Gamma\left(\frac{\lambda}{2}-\frac{i\omega}{2\pi T}\right)\Gamma\left(\frac{\lambda}{2}+\frac{i\omega}{2\pi T}\right)}{\Gamma(\lambda)},
\label{vsi}
\end{equation}
$\tau_0 \sim 1/\Delta_0$ is the QPS core size in time and $\Gamma (x)$ is the
Euler Gamma-function. Note that the result (\ref{Sa3}) turns out to be two times smaller as compared that derived within another measurement scheme \cite{SZ16a,SZ16b} in the corresponding limit. In other words, shot noise measured by each of our two detectors is 4 times smaller than that detected with the aid of the setup \cite{SZ16a,SZ16b}. The result (\ref{Sa3}) is illustrated in Fig. 2.
\begin{figure}
  \includegraphics[width=\columnwidth]{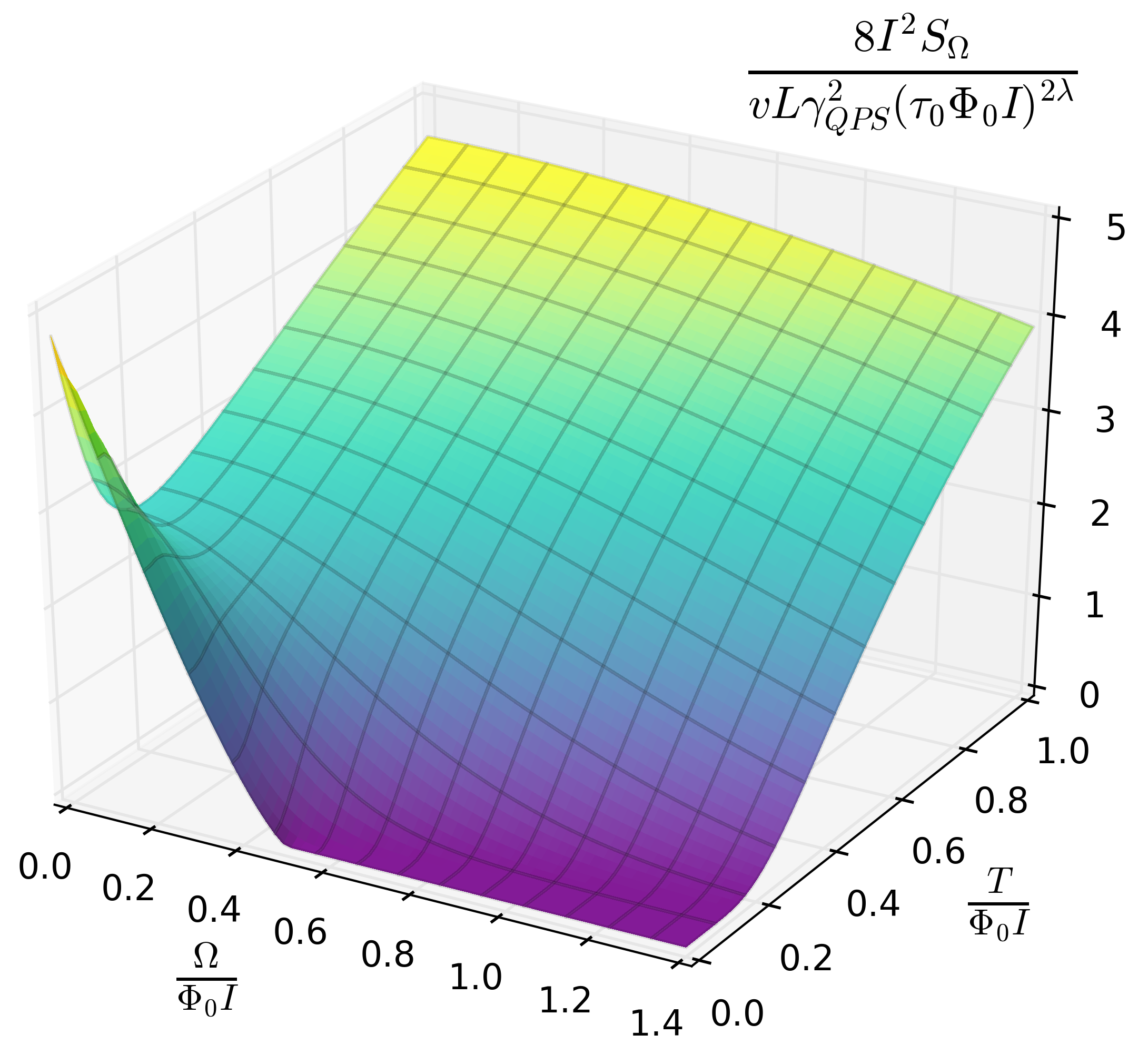}
  \caption{The dependence of QPS noise power $S_\Omega$ (\ref{Sa3}) at $\lambda=2.3$ from frequency $\Omega$ and temperature $T$. }
\end{figure}

At $T\to 0$ from Eq. (\ref{Sa3}) we find
\begin{equation}
S_{\Omega}^{QPS}\propto
\begin{cases}
I^{\lambda -1}(I-2\Omega /\Phi_0)^{\lambda -1}, & \Omega < \Phi_0I/2,
\\
0, & \Omega > \Phi_0I/2.
\end{cases}
\label{Sa4}
\end{equation}
In order to interpret this threshold behaviour let us bear in mind that 
at $T=0$ each QPS event can excite 2$N$ plasmons ($N=1,2...$) with total energy $E=\Phi_0 I$ and total zero momentum. The left and the right moving plasmons (each group carrying total energy $E/2$) eventually reach respectively the left and the right voltage probes which then detect voltage fluctuations with frequency $\Omega$. As in the long wire limit and for non-zero $\Omega$ these two groups of plasmons become totally uncorrelated, it is obvious that at $T=0$ voltage noise can only be detected at $\Omega < E/2$ in the agreement with Eq. (\ref{Sa4}).

In summary, we investigated QPS-induced voltage fluctuations in superconducting nanowires within the four-point measurement scheme. We demonstrated that shot noise detected in such nanowires may essentially
depend on the particular measurement setup. In long wires and at non-zero frequencies quantum voltage noise
is essentially determined by plasmons created by QPS events and propagating in opposite directions along the wire. As a result, at $T \to 0$ the shot noise vanishes at frequencies exceeding $\Phi_0 I/2$.

This work was supported in part by RFBR grant No. 15-02-08273. It has been presented at the International Conference SUPERSTRIPES 2016 \cite{superstr20016}.

\end{document}